\documentclass[twocolumn,showpacs,prb, longbibliography]{revtex4-1}%
\usepackage{amsfonts}
\usepackage{amsmath}
\usepackage{amssymb}
\usepackage{subfigure}
\usepackage{graphicx}
\usepackage{txfonts}
\usepackage{epstopdf}
\usepackage{makecell}
\usepackage[mathlines]{lineno}
\usepackage[colorlinks,linkcolor=blue,citecolor=blue,urlcolor=blue]{hyperref}%
\providecommand{\U}[1]{\protect\rule{.1in}{.1in}}

\begin{document}

\title{Spin-polarized scanning tunneling microscopy measurement scheme for determining the quantum geometric tensor}
\author{Shu-Hui Zhang$^{1}$}
\email{shuhuizhang@nankai.edu.cn}
\author{Jin Yang$^{2}$}
\author{Ding-Fu Shao$^{3}$}
\author{Jia-Ji Zhu$^{4}$}
\author{Wen-Long You$^{5}$}
\author{Wen Yang$^{2}$}
\email{wenyang@csrc.ac.cn}
\author{Kai Chang$^{6,7}$}
\email{kchang@zju.edu.cn}

\affiliation{$^{1}$School of Physics, Nankai University, Tianjin 300071, China}
\affiliation{$^{2}$Beijing Computational Science Research Center, Beijing 100193, China}
\affiliation{$^{3}$Key Laboratory of Materials Physics, Institute of Solid State Physics, HFIPS, Chinese Academy of Sciences,
Hefei 230031, China}
\affiliation{$^{4}$School of Science, Chongqing University of Posts and Telecommunications, Chongqing 400061, China}
\affiliation{$^{5}$College of Physics, Nanjing University of Aeronautics and Astronautics, Nanjing 211106, China}
\affiliation{$^{6}$Center for Quantum Matter, School of Physics, Zhejiang University, Hangzhou 310027, China}
\affiliation{$^{7}$Institute for Advanced Study in Physics, Zhejiang University, Hangzhou 310027, China}

\begin{abstract}
The quantum geometric tensor (QGT) embodies the geometry of the eigenstates of a
system's Hamiltonian, and its full characterization across diverse quantum
systems is essential. However, it is challenging to characterize the QGT of solid-state systems. Here we present an electric scheme to measure the
complete QGT of two-dimensional solid-state systems by using spin-polarized
scanning tunneling microscopy (STM), in which the spin texture is extracted from geometric amplitudes of
Friedel oscillations induced by the intentionally introduced magnetic impurity,
and then the QGT is derived from the momentum differential of spin texture. As
a canonical spin model, the surface states of a topological insulator offer a
promising way to demonstrate the scheme. In a slab of topological insulator, the
gapped surface states host complete QGT, i.e., nonvanishing quantum metric and
Berry curvature as its symmetric real part and the antisymmetric imaginary
part. Thus, a detailed derivation guides the use of the developed scheme to
measure the QGT of gapped surface states, even with an external magnetic
field. This study opens a new avenue to directly measure the complete QGT of
two-dimensional solid-state systems by using spin-polarized STM.

\end{abstract}
\maketitle

\section{Introduction}

Geometric understanding of various physical systems is remolding modern
physics\cite{PhysRevLett.131.240001}. For a quantum system, there is a
Hamiltonian whose solution gives the energy dispersion and eigenstates. The
geometric properties of eigenstates are encoded in a quantum geometric tensor
(QGT) including the well-established Berry
curvature\cite{rspa.1984.0023,s42254-019-0071-1,RevModPhys.82.1959} and the
developing quantum
metric\cite{Provost1980a,PhysRevE.76.022101,Kolodrubetz2017}. The Berry
curvature, as the antisymmetric part of the QGT, is well known due to its
significant role in topological
invariants\cite{RevModPhys.82.3045,RevModPhys.83.1057}, the anomalous Hall
effect\cite{RevModPhys.82.1539}, the spin Hall effect\cite{RevModPhys.87.1213},
the valley Hall effect\cite{PhysRevLett.99.236809,PhysRevLett.108.196802}, etc.
Conversely, the quantum metric, as the symmetric part of the QGT, is garnering
increasing interest due to its key role in orbital magnetic
susceptibility\cite{PhysRevB.91.214405,PhysRevB.94.134423}, the exciton Lamb shift
in transition-metal dichalcogenides\cite{PhysRevLett.115.166802}, the intrinsic
nonlinear Hall
effect\cite{PhysRevLett.112.166601,science.adf1506,s41586-023-06363-3},
electron-phonon coupling\cite{s41567-024-02486-0}, etc. In light of the demand
for complete local information in momentum space, there is strong enthusiasm
in performing the separated measurement of Berry
curvature\cite{science.aad4568,nphys4050,s41467-018-03397-4,aay2730,PhysRevLett.126.256601,PhysRevLett.131.133601,PhysRevLett.131.196603}
or the quantum metric\cite{PhysRevLett.127.107402,PhysRevLett.122.210401}, and
even the full measurement of
the QGT\cite{s41586-020-1989-2,nwz193,PhysRevLett.124.197002,s41467-020-20845-2,PhysRevLett.121.020401,PhysRevLett.131.156901,s41567-024-02678-8}%
. Different schemes have been proposed to measure the QGT of various systems
such as photonic lattices\cite{s41586-020-1989-2,s41467-020-20845-2}, coupled
qubits in diamond\cite{nwz193}, and Josephson
junctions\cite{PhysRevLett.124.197002}, yet the appropriate scheme for solid state systems is less
explored\cite{PhysRevLett.131.156901,s41567-024-02678-8,ado6049}.

Recently, the complete QGT of exciton-photon modes was directly measured in a
two-dimensional high-finesse planar microcavity through the Stokes vectors
determined by the polarization intensities of
photoluminescence\cite{s41586-020-1989-2}. Specifically, the QGT components
can be given through the momentum differential of Stokes vectors:
\begin{subequations}
\label{NQGT}%
\begin{align}
g_{ij}  &  =\frac{1}{4}(\partial_{k_{i}}\theta\partial_{k_{j}}\theta+\sin
^{2}\theta\partial_{k_{i}}\phi\partial_{k_{j}}\phi),\\
\Omega_{z}  &  =\frac{1}{2}\sin\theta(\partial_{_{k_{i}}}\theta\partial
_{k_{j}}\phi-\partial_{k_{j}}\theta\partial_{k_{i}}\phi),
\end{align}
\end{subequations}
for an eigenstate $\left\vert u(\mathbf{k})\right\rangle =\left[
\cos({\theta}/{2}),
\sin({\theta}/{2})e^{i\phi}
\right]^\text{T} $
with the polar angle $\theta$ and azimuth $\phi$. Here, $g_{ij}$ and
$\Omega_{z}$ are, respectively, the components of the quantum metric and Berry
curvature for the two-dimensional system. In view of the close analogy between the
Stokes vector of light and the spin vector of the electron, the QGT of the
solid state system is expected to be given by measuring the momentum-space distribution of 
the spin vector, i.e., spin texture.

The measurement of spin texture bears its own
importance\cite{nature08308,science.1167733}. For example, for the chiral spin
texture of surface states of a topological insulator (TISS), experiments demonstrated its protection from
backscattering\cite{nature08308,science.1167733} and its evolution with the
quantum tunneling coupling in slabs\cite{PhysRevLett.112.057601,ncomms4841},
hexagonal warping degree\cite{PhysRevLett.106.216803} and magnetic
doping\cite{nphys2351}. The main technology for measuring the spin texture is
limited to spin- and angle-resolved photoemission spectroscopy. Very recently,
an experiment\cite{ado6049} used photoemission spectroscopy to directly
measure the quantum metric in solids according to Eq. (\ref{NQGT}). However,
the optical measurement of spin texture in the experiment\cite{ado6049} was
limited to the $\theta=0$ case, which can not consistently provide the
complete QGT. Therefore, the general measurement of spin texture is highly anticipated.

\begin{figure*}[ptbh]
\includegraphics[width=2.0\columnwidth,clip]{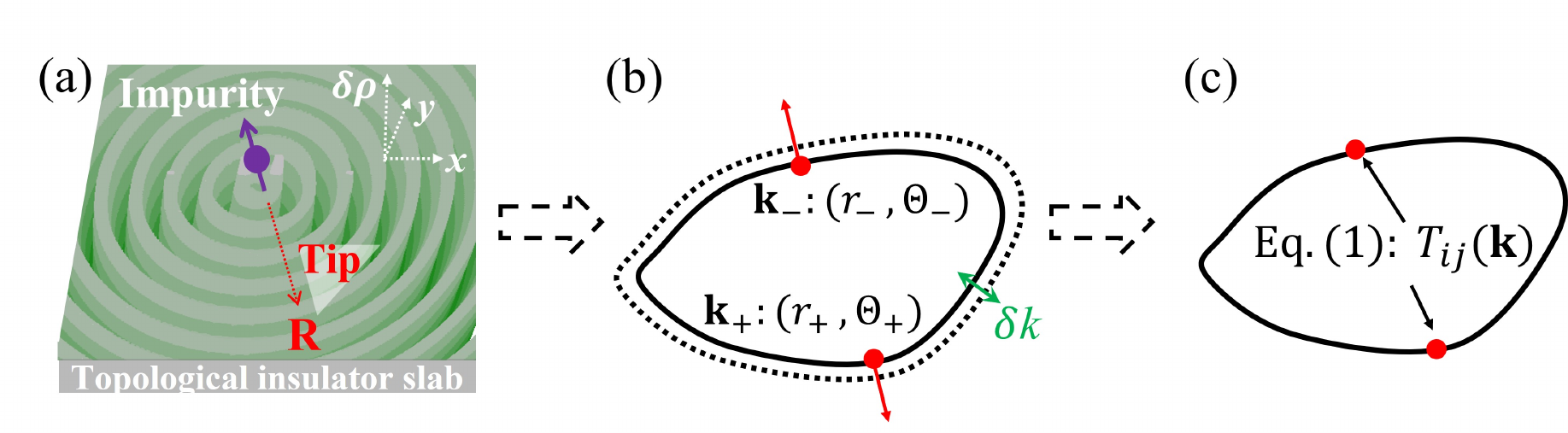}\caption{Schematic STM measurement
scheme for the quantum geometric tensor. (a) In a topological insulator slab,
spin-polarized Friedel oscillations are induced by the intentionally
introduced magnetic impurity, which can be mapped out by the spin-polarized
STM measurement. (b) Along any direction $\mathbf{R}$ in real space, the
long-range Friedel oscillations are mainly contributed by the coupled
backscattering states (red dots) on the constant-energy contour with group velocities (red
arrows) parallel to ${\pm\mathbf{R}}$. For coupled backscattering states at
$\mathbf{k}_{\pm}$, the spin vectors characterized by ($r_{\pm}, \Theta_{\pm}%
$) can be extracted from the geometric amplitudes of Friedel oscillations,
which form the spin texture on the constant-energy contour. Based the
energy-resolved STM measurement, the spin textures on different
constant-energy (solid and dashed) contours should be provided. (c) If the STM
measurement has sufficiently high momentum resolution for spin texture, i.e.,
sufficiently small $\delta k$ in (b) depending on the energy resolution of STM
measurement, the quantum geometric tensor can be given through the momentum
differential of spin texture, i.e., Eq. (\ref{NQGT}). }%
\label{scheme}%
\end{figure*}

Beyond photoemission spectroscopy, scanning tunneling microscopy (STM) is
a powerful electric tool for characterizing the electronic structure, with
atomic-scale spatial resolution, high energy resolution, and compatibility
with the magnetic field\cite{s42254-021-00293-7}. To our knowledge, STM has
not yet been able to measure the spin texture. The spin information manifests
itself in the geometric amplitudes of spin-polarized Friedel oscillations
(FOs)\cite{PhysRevLett.133.036204} measurable by STM; this echoes the
measurement of QGT through the Stokes vector\cite{s41586-020-1989-2}. In this
work, we develop an electric scheme to measure the general spin texture
($\theta\neq0$) and then QGT by using spin-polarized STM. For the sake of
universality, the gapped TISS [Fig. \ref{scheme}(a)] of the paradigmatic spin
model with a complete QGT is used to demonstrate the scheme. This scheme first
extracts the high momentum-resolved spin texture [Fig. \ref{scheme}(b)]
from geometric amplitudes of FOs induced by the intentionally introduced
magnetic impurity, and then derives the QGT [Fig. \ref{scheme}(c)] from
the momentum differential of spin texture according to Eq. (\ref{NQGT}). This
study provides an electric scheme for directly measuring the complete QGT of
two-dimensional solid-state systems, and hints at the great potential of the
information extraction from the geometric amplitudes of STM.

\section{Model and measurable QGT}

To consider a model spin system, i.e., TISS, the general Hamiltonian of the
Dirac fermions is given by (see Refs. \onlinecite{PhysRevLett.132.096302} and
Supplementary Material\cite{2024-SM})
\begin{equation}
\hat{H}=k_{x}\sigma_{y}-k_{y}\sigma_{x}+\delta\sigma_{z}+\mathbf{t\cdot k}.
\label{RHF}%
\end{equation}
Here, $\hbar v_{F}\equiv1$ is used for dimensionless derivations with $v_{F}$
for the Fermi velocity, $\delta$ denotes the gap opening (e.g., due to the
intersurface coupling in a slab\cite{PhysRevB.81.115407} or an out-of-plane
magnetic field\cite{PhysRevB.87.085431}), and $\mathbf{t}$ represents the tilt
velocity vector (e.g., due to the in-plane magnetic
field\cite{PhysRevB.101.041408}). $\sigma_{x,y,z}$ are Pauli matrices for
the spin vectors. To solve the Hamiltonian, one can obtain the energy
dispersion and wave function, respectively, $E_{\mathbf{k}}=\mathbf{t\cdot
k}+\sqrt{k^{2}+\delta^{2}}$ and $|\psi_{\mathbf{k}}\rangle=|u(\mathbf{k}%
)\rangle e^{i\mathbf{k\cdot R}}$ (for the conduction band without loss of
generality) in which the spinor part is
\begin{equation}
|u(\mathbf{k})\rangle=\frac{1}{\sqrt{1+r^{2}}}\left[
\begin{array}
[c]{c}%
1\\
re^{i\Theta}%
\end{array}
\right]  \label{WF2}%
\end{equation}
with the spinor component $re^{i\Theta}=(-k_{y}+ik_{x})/(\sqrt{k^{2}%
+\delta^{2}}+\delta)$. We note here that the expression of Eq. (\ref{WF2}) is
crucial for generalizing the STM measurement of spin texture from the gapless
case\cite{PhysRevLett.133.036204} to the gapped case\cite{2024-SM} and then
arriving at the scheme for QGT. To relate two spinor expressions for Stokes
vector and TISS to each other, we have $e^{i\theta}=(1-r^{2},2ir)/(1+r^{2})$
and $\phi=\Theta$, which are helpful for introducing the spin vector as
\begin{subequations}
\label{SV}%
\begin{align}
\mathbf{S}  &  \equiv\left\langle u(\mathbf{k})\right\vert (\sigma_{x}%
,\sigma_{y},\sigma_{z})\left\vert u(\mathbf{k})\right\rangle ,\\
&  =(\frac{2r\cos\Theta}{1+r^{2}},\frac{2r\sin\Theta}{1+r^{2}},\frac{1-r^{2}%
}{1+r^{2}}).
\end{align}
This corroborates the availability of Eq. (\ref{NQGT}) for the extraction of
QGT as shown later.

Usually, by using the Bloch states $\left\vert u_{n}\right\rangle $ in momentum
space, the QGT is defined
as\cite{PhysRevLett.131.156901,s41567-024-02678-8,ado6049}
\end{subequations}
\begin{equation}
T_{ij}^{n}=\left\langle \mathbf{\partial}_{k_{i}}u_{n}|\mathbf{\partial
}_{k_{j}}u_{n}\right\rangle -\left\langle \mathbf{\partial}_{k_{i}}u_{n}%
|u_{n}\right\rangle \left\langle u_{n}|\mathbf{\partial}_{k_{j}}%
u_{n}\right\rangle . \label{QGT}%
\end{equation}
Its symmetric real and antisymmetric imaginary parts correspond to quantum
metric $g_{ij}^{n}=\operatorname{Re}[T_{ij}^{n}]$\ and Berry curvature
$\Omega_{ij}^{n}=-2\operatorname{Im}[T_{ij}^{n}]$, respectively. The presence
or absence of Berry curvature and quantum metric depends on the symmetry of
the Hamiltonian; e.g., $PT$ symmetry usually enforces vanishing Berry
curvature except at some singular points. For the conduction band of the
Hamiltonian of Eq. (\ref{RHF}), the QGT includes four independent
components\cite{PhysRevLett.132.096302}, which are (omitting of the band index
$n$)
\begin{subequations}
\label{AQGT}%
\begin{align}
\Omega_{z}  &  =-\frac{\delta}{2(k^{2}+\delta^{2})^{3/2}},\\
g_{xx}  &  =\frac{\delta^{2}+k_{y}^{2}}{4(\delta^{2}+k^{2})^{2}},\\
g_{yy}  &  =\frac{\delta^{2}+k_{x}^{2}}{4(\delta^{2}+k^{2})^{2}},\\
g_{xy}  &  =-\frac{k_{x}k_{y}}{4(\delta^{2}+k^{2})^{2}}.
\end{align}

The quantum metric and Berry curvature are both gauge-invariant geometric
quantities for the Bloch states under transformation $\left\vert
u_{n}\right\rangle \rightarrow e^{i\varphi}\left\vert u_{n}\right\rangle $
with $e^{i\varphi}$ an arbitrary phase factor. While Berry curvature is known for its importance in comprehending various anomalous Hall
effects\cite{rspa.1984.0023,s42254-019-0071-1,RevModPhys.82.1959}, the quantum
metric quantifies the geometric distance between neighboring quantum states
and has only recently gained considerable
attention\cite{Provost1980a,Kolodrubetz2017}. According to Eq. (\ref{QGT}),
the QGT can be formulated if the Bloch states can be identified. Usually, it
is a difficult task to identify the Bloch states of a multiband system. But
for the two-band system, up to a global undetermined phase factor
$e^{i\varphi}$, the Bloch states $\left\vert u_{n}\right\rangle $\ can be
characterized by the spin vector [Eq. (\ref{SV})] since it is
gauge invariant, which then brings about the applicability of Eq. (\ref{NQGT}) for
QGT. In the following, we develop a feasible scheme to measure the spin vector
of $\left\vert u_{n}\right\rangle $ by using spin-polarized STM measurement of
FOs induced by the designed single magnetic impurity, and then to characterize
the QGT through the momentum differential of spin texture.

\section{Results and discussions}

Here, we first introduce the spin-polarized STM scheme to measure the QGT,
with a focus on its underlying physics, specific applications and typical
features. In light of the paradigmatic role and intensive studies, the gapped
TISS is used to demonstrate the proposed scheme, in which both isotropic and
tilted cases are considered. In particular, the isotropic case is expected to
be verified in the near future, and the tilted case is induced by an in-plane
magnetic field\cite{PhysRevB.101.041408} which further shows the superiority of STM over the photoemission spectroscopy\cite{ado6049}.

\subsection{STM measurement scheme}

Using the standard $T$-matrix
approach\cite{economou2006green,PhysRevLett.133.036204}, one can obtain FOs
characterized by the change of the local density of
states\cite{PhysRevB.81.233405,PhysRevB.85.125314}:
\end{subequations}
\begin{equation}
\delta\rho_{\alpha\beta}\left(  \varepsilon,\mathbf{R}\right)  =-\frac{1}{\pi
}\operatorname{Im}\mathrm{\operatorname*{Tr}}[\mathbf{g(}\varepsilon
,\mathbf{R)T}_{\alpha}\mathbf{g(}\varepsilon,\mathbf{-R)}\sigma_{\beta
}\mathbf{]},\label{LDOS}%
\end{equation}
where the $T$ matrix is expressed as%
\begin{equation}
\mathbf{T}_{\alpha}\left(  \varepsilon\right)  =\mathbf{V}_{\alpha}\left[
1-\mathbf{g}\left(  \varepsilon,\mathbf{0}\right)  \mathbf{V}_{\alpha}\right]
^{-1},\label{Tmatrix}%
\end{equation}
with $\mathbf{V}_{\alpha}$ representing the magnetic impurity potential. Here,
we use the subscript $\alpha\in\{0,x,y,z\}$ with $\alpha=0$ and $\alpha\neq0$
for the spin-unpolarized and spin-polarized impurities or STM tip, respectively,
so $\delta\rho_{\alpha\beta}\left(  \varepsilon,\mathbf{r}\right)  $ gives
$\beta$-resolved FOs induced by the local $\alpha$-resolved impurity $\mathbf{V}%
_{\alpha}=V_{0}\sigma_{\alpha}$ of strength $V_{0}$. $\mathbf{g(}%
\varepsilon,\mathbf{R)}$ is the bare Green's function (GF) of the host system for
the impurity.

The underlying physics of the proposed scheme can be explained through Fig.
\ref{scheme}. In Fig. \ref{scheme}(a), FOs described by Eq. (\ref{LDOS}) are
quantum interference of two propagation waves: one is the propagation
$\mathbf{g}(\varepsilon,\mathbf{-R})$ of emitted electron waves from the STM
tip $\sigma_{\beta}$ to the impurity, and the other is the propagation
$\mathbf{g}(\varepsilon,\mathbf{R})$ of scattered electron waves by the
impurity $\mathbf{T}_{\alpha}$ back to the STM tip. So FOs embody the
information of TISS and impurity. If the impurity is designed, e.g., with weak
strength and specific polarized direction, one obtains $\mathbf{T}_{\alpha
}\simeq\mathbf{V}_{\alpha}$. Then, the impurity can be regarded as the other
tip, and FOs are the linear responses of TISS measured by two atomic-scale
leads\cite{PhysRevLett.133.036204}. In particular, long-range GFs have the matrix form 
$\mathbf{g}(\varepsilon,\pm\mathbf{R})\propto\left\vert u(\mathbf{k}_{\pm
})\right\rangle \left\langle u(\mathbf{k}_{\pm})\right\vert $, in which
$\mathbf{k}_{\pm}$ are the classical momenta of the coupled backscattering
states with group velocities parallel to $\pm\mathbf{R}$. The matrix forms of
GFs combine the geometric density of states to determine the (geometric)
amplitudes of FOs\cite{PhysRevLett.133.036204}. Thus, it is an inverse problem
to solve $\left\vert u(\mathbf{k}_{\pm})\right\rangle $ and/or the spin
vectors characterized by ($r_{\pm},\Theta_{\pm}$) from the geometric
amplitudes of FOs along any $\mathbf{R}$; see Eq. (\ref{WF2}). For any
constant-energy contour, the high momentum-resolved spin texture formed by the
spin vectors can be given by the atomic-scale resolved STM measurement. Then,
the high energy-resolved STM measurement\cite{Machida2018, s42005-023-01201-4}
favors the high momentum-resolved spin texture in the whole momentum space
[Fig. \ref{scheme}(b)], whose differential gives the QGT [Fig.
\ref{scheme}(c)].

Specific to the system of TISS, impurity-induced FOs can be rewritten into the form:%
\begin{equation}
\delta\rho_{\alpha\beta}=\delta\rho_{a,\alpha\beta}\cos[(k_{+}+k_{-}%
)R]+\delta\rho_{b,\alpha\beta}\sin[(k_{+}+k_{-})R],\label{UFO}%
\end{equation}
whose amplitudes $\delta\rho_{a/b,\alpha\beta}$ can be extracted from
the calculations of the exact $T$-matrix approach or experimental results. If
$\delta\rho_{a/b,\alpha\beta}$ satisfy the Born approximation, their
spin-resolved components have the analytical expressions ($\alpha\neq0$ and
$\beta\neq0$)\cite{2024-SM}
\begin{subequations}
\label{TLDOS}%
\begin{align}
\delta\rho_{a,xx} &  =\mathcal{C}\left[  r_{+}^{2}+r_{-}^{2}+2r_{+}r_{-}%
\cos\left(  \Theta_{-}+\Theta_{+}\right)  \right]  ,\\
\delta\rho_{a,yy} &  =\mathcal{C}\left[  r_{+}^{2}+r_{-}^{2}-2r_{+}r_{-}%
\cos\left(  \Theta_{-}+\Theta_{+}\right)  \right]  ,\\
\delta\rho_{a,zz} &  =\mathcal{C}\left[  1+r_{+}^{2}r_{-}^{2}-2r_{+}r_{-}%
\cos\left(  \Theta_{-}-\Theta_{+}\right)  \right]  ,\\
\delta\rho_{a,xy} &  =\mathcal{C}2r_{+}r_{-}\sin\left(  \Theta_{+}+\Theta
_{-}\right)  ,\\
\delta\rho_{b,xy} &  =\mathcal{C}\left(  r_{+}^{2}-r_{-}^{2}\right)  ,\\
\delta\rho_{a,xz} &  =\mathcal{C}\left[  r_{-}\left(  1-r_{+}^{2}\right)
\cos\Theta_{-}+r_{+}\left(  1-r_{-}^{2}\right)  \cos\Theta_{+}\right]  ,\\
\delta\rho_{b,xz} &  =-\mathcal{C}\left[  r_{-}\left(  1+r_{+}^{2}\right)
\sin\Theta_{-}-r_{+}\left(  1+r_{-}^{2}\right)  \sin\Theta_{+}\right]  ,\\
\delta\rho_{a,yz} &  =\mathcal{C}\left[  r_{-}\left(  1-r_{+}^{2}\right)
\sin\Theta_{-}+r_{+}\left(  1-r_{-}^{2}\right)  \sin\Theta_{+}\right]  ,\\
\delta\rho_{b,yz} &  =\mathcal{C}\left[  r_{-}\left(  1+r_{+}^{2}\right)
\cos\Theta_{-}-r_{+}\left(  1+r_{-}^{2}\right)  \cos\Theta_{+}\right]  ,
\end{align}
which represent geometric amplitudes inheriting from the geometric density of
states\cite{PhysRevLett.133.036204}. Thus, it is an inverse problem to find
$\mathcal{C}$, $r_{\pm}$, and $\Theta_{\pm}$ from $\delta\rho_{a/b,\alpha
\beta}$, which are solved analytically for TISS\cite{2024-SM}: %

\end{subequations}
\begin{subequations}
\label{RPM}%
\begin{align}
\mathcal{C} &  =\frac{1}{4}\left(  c_{0}-\sqrt{\Lambda+c_{0}\delta\rho_{a,zz}%
}\right)  ,\\
r_{\pm} &  =\sqrt{\frac{\left(  \delta\rho_{a,xx}+\delta\rho_{a,yy}\right)
}{4\mathcal{C}}\pm\frac{\delta\rho_{b,xy}}{2\mathcal{C}}},\\
\sin\Theta_{\pm} &  =\frac{r_{\pm}^{2}\left(  \delta\rho_{a,yz}\mp\delta
\rho_{b,xz}\right)  +\left(  \delta\rho_{a,yz}\pm\delta\rho_{b,xz}\right)
}{2\mathcal{C}r_{\pm}(1-r_{+}^{2}r_{-}^{2})},\\
\cos\Theta_{\pm} &  =\frac{r_{\pm}^{2}\left(  \delta\rho_{a,xz}\pm\delta
\rho_{b,yz}\right)  +\left(  \delta\rho_{a,xz}\mp\delta\rho_{b,yz}\right)
}{2\mathcal{C}r_{\pm}(1-r_{+}^{2}r_{-}^{2})},
\end{align}
with
\end{subequations}
\begin{subequations}
\begin{align}
c_{0} &  \equiv2\delta\rho_{a,zz}-2\sqrt{\Lambda_{0}+\Lambda+\delta\rho
_{a,zz}^{2}},\\
\Lambda_{0} &  \equiv\left(  \delta\rho_{a,xx}+\delta\rho_{a,yy}\right)
^{2}-4\left(  \delta\rho_{b,xy}\right)  ^{2},\\
\Lambda &  \equiv2(\delta\rho_{a,xz}^{2}+\delta\rho_{a,yz}^{2})-2(\delta
\rho_{b,xz}^{2}+\delta\rho_{b,yz}^{2}).
\end{align}
\end{subequations}
As a result, the spin vectors characterized by ($r_{\pm},\Theta_{\pm}$) are
given from the geometric amplitudes of FOs along any $\mathbf{R}$\textbf{. }

\begin{figure}[ptbh]
\includegraphics[width=1.0\columnwidth,clip]{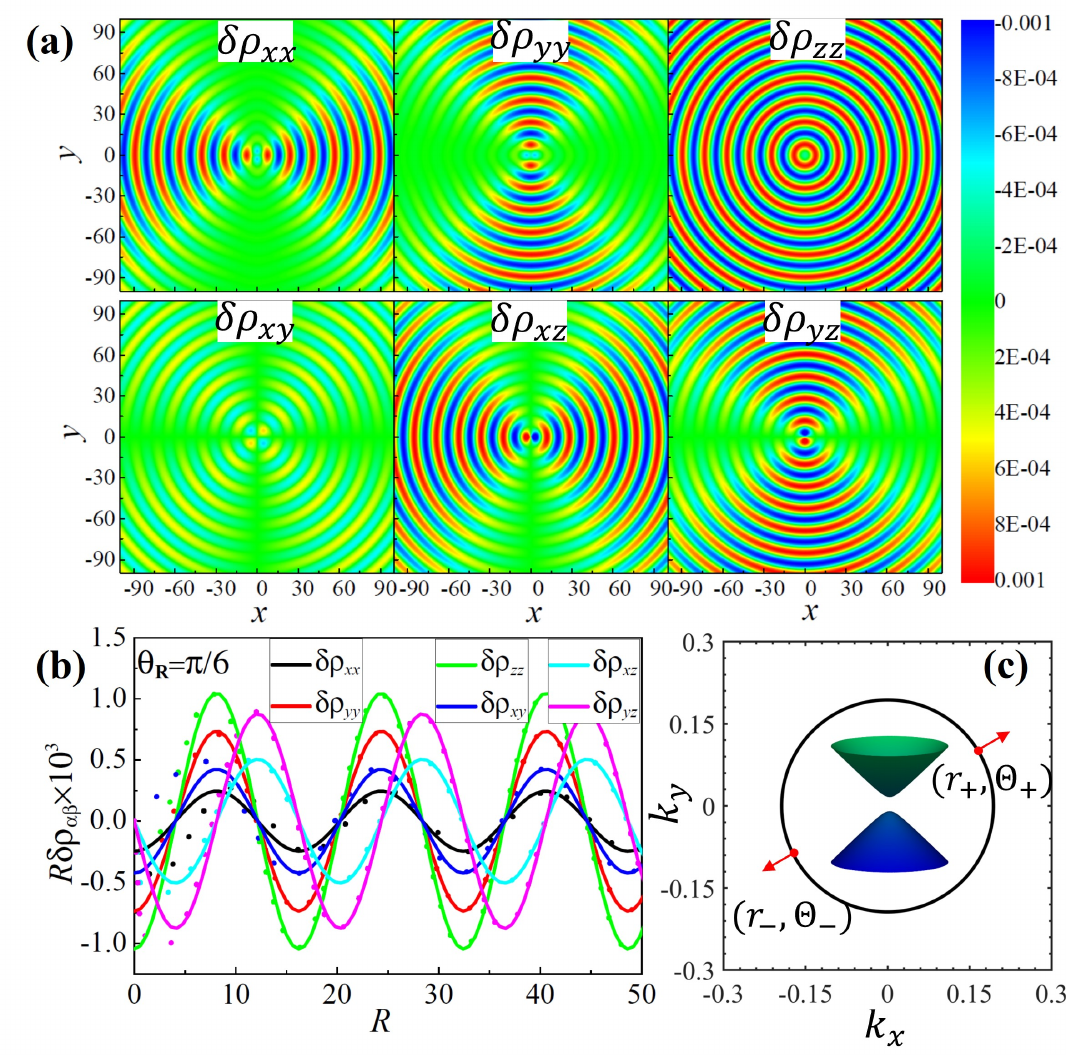}\caption{Spin-polarized
Friedel oscillations induced by a single magnetic impurity for isotropic
gapped TISS. (a) $\delta\rho_{\alpha\beta}$ calculated by the $T$-matrix approach.
(b) Comparison of $\delta\rho_{\alpha\beta}$ from the $T$-matrix approach (dotted
lines) and the Born approximation (solid lines). (c) For the TISS with isotropic
gapped Dirac cone (inset), the constant-energy contour is a circle. The
amplitudes of real-space Friedel oscillations along $\mathbf{R}$ in (b)
determine the momentum-space spin vectors ($r_{\pm}, \Theta_{\pm}$) of coupled
backscattering states (red dots) with group velocities (red arrows) parallel
to ${\pm\mathbf{R}}$. Here, $V_{0}=0.1$, $\varepsilon=0.2$, and $\delta=0.05$.
}%
\label{tiltless}
\end{figure}

For the general applications, the proposed scheme are summarized into three
key steps as shown by Fig. \ref{scheme}: (I) inputting $\delta\rho_{\alpha
\beta}\left(  \varepsilon,\mathbf{R}\right)  $ in real space from experiments
or simulations; (II) solving the spin vectors ($r_{\pm}, \Theta_{\pm}$) through
the geometric amplitudes of $\delta\rho_{\alpha\beta}\left(  \varepsilon
,\mathbf{R}\right)  $ for any $\mathbf{R}$\cite{2024-SM}, which form the high
momentum-resolved spin texture on the constant-energy contour $E_{\mathbf{k}%
}=\varepsilon$ due to the atomic-scale spatial resolution of STM; (III)
deriving QGT through the momentum differential as the difference of spin
textures on two constant-energy contours $E_{\mathbf{k}}=\varepsilon$ and
$E_{\mathbf{k}}^{\prime}=\varepsilon+\delta\varepsilon$ with the later given
by repeating (I) and (II). STM has the high energy resolution, namely the
small $\delta\varepsilon$, that ensures the momentum differential.

The proposed general scheme for the two-band model has two distinct features, i.e., realistic measurement by an electric means
and the compatibility with magnetic field. Instead of the developed optical
method\cite{s41567-024-02678-8,ado6049}, the proposed scheme realizes an
electric probe of the QGT of solids systems. Subsequently, we demonstrate this
scheme through gapped TISS in the absence and presence of the tilt induced by
an in-plane magnetic field. Due to intensive
theoretical\cite{PhysRevB.80.245317,PhysRevB.82.155142,PhysRevB.85.125314,PhysRevLett.102.156603,PhysRevLett.106.097201,PhysRevLett.106.136802,PhysRevLett.107.076801,PhysRevLett.109.266405,PhysRevB.86.165313,PhysRevB.89.195417,PhysRevB.95.115102,PhysRevB.105.075306}
and experimental
studies\cite{nature08308,ncomms6349,PhysRevB.85.081305,PhysRevB.85.205317,PhysRevLett.103.266803,PhysRevLett.106.166805,PhysRevLett.106.206805,PhysRevLett.107.056803,PhysRevLett.108.117601,PhysRevLett.108.256810,PhysRevLett.108.256811,PhysRevLett.110.126804,PhysRevLett.111.176802}%
, the gapped TISS is the most promising platform to verify our scheme in the
near future. In particular, probing the spin texture and/or QGT of interesting
tilted
TISS\cite{PhysRevLett.115.216806,PhysRevB.108.035407,PhysRevLett.131.246301}
is beyond the ability of optical method due to its incompatibility with
magnetic field\cite{s42254-021-00293-7}. We note that the proposed scheme also
provides high momentum-resolved measurement of spin texture, which is
intrinsically an important issue for STM technology. In addition, we later
employ an analytical scheme and a numerical scheme for the amplitude
extraction in step (II) through analytical Born approximation and numerical
fitting of $\delta\rho_{\alpha\beta}\left(  \varepsilon,\mathbf{R}\right)
$\cite{2024-SM}, respectively.

\begin{figure}[ptbh]
\includegraphics[width=1.0\columnwidth,clip]{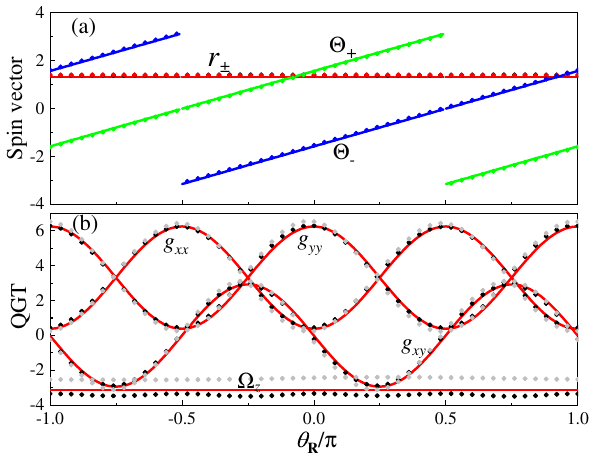}\caption{Spin
vector and quantum geometric tensor for gapped TISS. (a) The spin vectors
characterized by ($r_{\pm}, \Theta_{\pm}$) for the coupled backscattering
states. To consider $V_{0}=0.1$, the results are from the analytical scheme
(solid lines) and the numerical scheme (circles). (b) The quantum geometric
tensor including four independent components such as $\Omega_{z}$, $g_{xx}$,
$g_{yy}$, $g_{xy}$ from the numerical scheme (light grey and black circles)
and the analytical scheme (red lines), and $V_{0}=0.1$ is for red and light
grey lines while $V_{0}=10^{-3}$ for black ones. Here, $\varepsilon=0.2$,
$\delta=0.05$, and $\delta k=10^{-5}$. }%
\label{Qtiltless}
\end{figure}

\subsection{Isotropic case}

Here, we use the proposed three-step scheme for the isotropic case of gapped
TISS, in step (I) $\delta\rho_{\alpha\beta}\left(  \varepsilon,\mathbf{R}%
\right)$ are given through simulations by using the exact $T$-matrix
approach, as shown by Fig. \ref{tiltless}(a). The well-known
dimension-determined $1/R$ decay for two-dimensional TISS is used to rescale
FOs, i.e., $R\delta\rho_{\alpha\beta}$. Although the Fermi surface of gapped
TISS is isotropic [Fig. \ref{tiltless}(c)], the spin-polarized FOs
$\delta\rho_{\alpha\beta}$ are typically anisotropic, as expected from the
analytical expressions under the Born approximation\cite{2024-SM}. In
particular, the numerical results from the $T$-matrix approach and those from the Born
approximation for $\delta\rho_{\alpha\beta}$ agree very well even when the
oscillations go beyond just one period, as demonstrated by Fig. \ref{tiltless}(b).

In step (II), the geometric amplitudes of FOs can be extracted for the coupled
backscattering states of Fig. \ref{tiltless}(c) with group velocities parallel
to $\pm\mathbf{R}$ of Fig. \ref{tiltless}(b). The spin vector ($r_{\pm},
\Theta_{\pm}$) should be solved from the geometric amplitudes of FOs, shown by
the solid lines in Fig. \ref{Qtiltless}(a) featured by $r_{+}=r_{-}$ and
$e^{i\Theta_{+}}=-e^{i\Theta_{-}}$ for isotropic gapped TISS [Eq.
(\ref{WF2})]. As a result, the spin vectors in momentum space of coupled
backscattering states are given by FOs measurement in real space. Using Fig.
\ref{tiltless}(a), the high momentum-resolved spin texture on the
constant-energy contour is obtained since STM has the atomic-scale spatial resolution.

\begin{figure}[ptbh]
\includegraphics[width=1.0\columnwidth,clip]{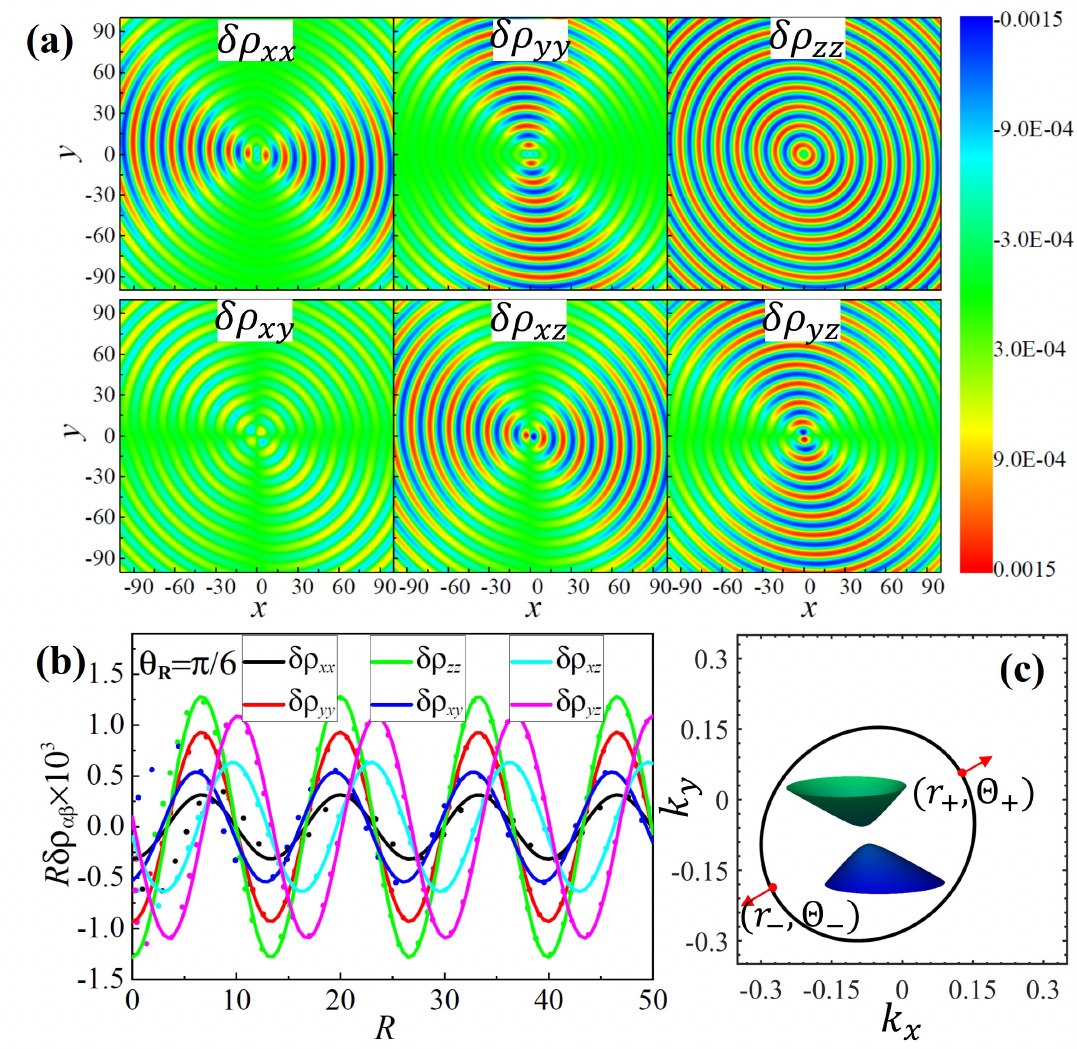}\caption{Spin-polarized
Friedel oscillations induced by a single magnetic impurity for tilted gapped
TISS. (a) $\delta\rho_{\alpha\beta}$ calculated by the $T$-matrix approach. (b)
Comparison of $\delta\rho_{\alpha\beta}$ from the $T$-matrix approach (dotted
lines) and the Born approximation (solid lines). (c) For the TISS with tilted
Dirac cone (inset), the constant-energy contour is an ellipse. The amplitudes
of real-space Friedel oscillations along $\mathbf{R}$ in (b) determine the
momentum-space spin vectors ($r_{\pm}, \Theta_{\pm}$) of the coupled
backscattering states (red dots) with group velocities (red arrows) parallel
to ${\pm\mathbf{R}}$. Here, $\mathbf{t}=(0.3,0.3)$, $V_{0}=0.1$,
$\varepsilon=0.2$, and $\delta=0.05$.}%
\label{tilt}
\end{figure}

\begin{figure}[ptbh]
\includegraphics[width=1.0\columnwidth,clip]{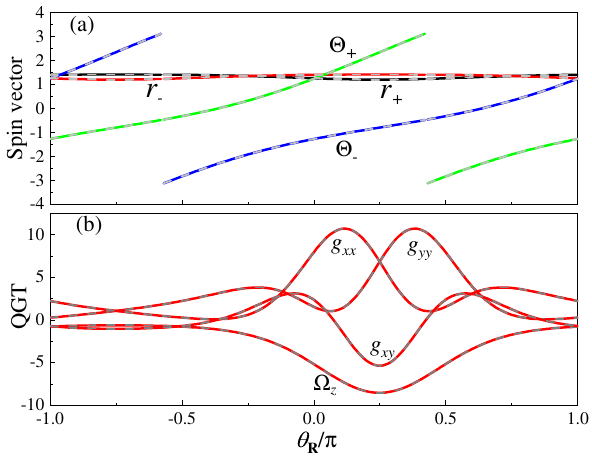}\caption{Spin vector
(a) and quantum geometric tensor (b) from the analytical scheme (solid lines)
and exact expressions (dashed lines) for tilted gapped TISS. Here,
$\mathbf{t}=(0.3,0.3)$, $V_{0}=0.1$, $\varepsilon=0.2$, $\delta=0.05$, and
$\delta k=10^{-5}$. }%
\label{Qtilt}
\end{figure}

In step (III), the QGT can be derived from the momentum differential of spin
texture. The high momentum-resolved spin textures for the proximate
constant-energy contours are realized through the high energy-resolved STM
measurement. Typically, the energy resolution of STM can be up to $10\mu eV$
at low temperature\cite{Machida2018, s42005-023-01201-4}. Adopting
$v_{F}=10^{6}$ cm/s and $\delta\varepsilon\propto\delta k$, a proper $\delta
k=10^{-5}$ is used in our calculations. In Fig. \ref{Qtiltless}(b), we note
that the analytical scheme is used here for QGT (red lines) and it always
reproduces the results from exact expressions (not shown for brevity), i.e.,
Eqs. (\ref{WF2}) and (\ref{AQGT}). We will discuss the impurity
strength effect by using a numerical scheme in the final section.

\subsection{Tilted case}

Recently, the tilted TISS induced by an in-plane magnetic field was proposed
to explain the planar Hall effect\cite{PhysRevB.101.041408}; it attracted wide
interest\cite{PhysRevB.103.155415,PhysRevLett.127.076601,PhysRevB.106.L081121,PhysRevB.107.245141,PhysRevLett.133.036204}%
. However, there is no experimental spectral evidence for the tilted TISS,
maybe due to the incompatibility of photoemission spectroscopy with the
magnetic field\cite{s42254-021-00293-7}. STM is a promising tool to characterize the
tilted TISS, in particular, their QGT, which is our focus. Following the three steps
of the scheme, Fig. \ref{tilt}(a) shows spin-polarized FOs for gapped TISS
with tilt, and displays the universal $1/R$ decay but becomes more
asymmetric compared to Fig. \ref{tiltless}(a). Similarly to Fig. \ref{tiltless}%
(b), $\delta\rho_{\alpha\beta}$ from the $T$-matrix approach tend towards to those
from the Born approximation by observing FOs beyond one period in Fig.
\ref{tilt}(b). Then, the geometric amplitude can be extracted for the coupled
backscattering states of Fig. \ref{tilt}(c) with group velocities parallel to
$\pm\mathbf{R}$ of Fig. \ref{tilt}(b). Figure \ref{Qtilt} shows the spin
texture and QGT by using the analytical scheme (solid lines) and the exact
expressions of Eqs. (\ref{WF2}) and (\ref{AQGT}) (dashed lines), which are
exactly the same. The numerical scheme is available for the small tilt
parameter (not shown) but not for the large tilt parameter, e.g.,
${\mathbf{t}}=(0.3,0.3)$, because of the intrinsic trouble in calculating the
on-site GF of a single Dirac
valley\cite{Wehling2009,PhysRevB.78.235438,PhysRevB.95.245137,PhysRevB.96.161113,PhysRevB.107.195409}
enforced by the Nielsen-Ninomiya theorem\cite{NIELSEN1983}. This trouble may be
tackled for a topological insulator slab by considering two Dirac valleys of
the lattice model in future\cite{PhysRevB.83.165425}.

\section{Outlook and conclusions}

The state of the art for spin-polarized STM technology
\cite{RevModPhys.75.933,MeierScience2008,ZhouNatPhys2010,science.1183224,AlexanderScience2011,science.aax8222,sciadv.abd7302}
favors the extraction of spin texture from FOs induced by a delicately designed
magnetic impurity, similar to the experiments for STM measurement of the Berry
phase\cite{s41586-019-1613-5,PhysRevLett.125.116804}. The proposed scheme for the QGT is based on the FOs in real space, which also can give the constant-energy
contour\cite{PhysRevLett.133.036204}, and thus bypasses the conventional
necessary Fourier transformation\cite{s42254-021-00293-7} and the concomitant
signal broadening due to finite image area. To realize the proposed scheme,
the magnetic impurity on TISS with magnetization directions $\sigma_{x,y,z}$
and weak $V_{0}$ is necessary. For the former, use of a magnetic field is promising for fixing the magnetization directions of the magnetic impurity, and its effect
on the electronic Hamiltonian of Eq. (\ref{RHF}) can be incorporated through
renormalizing the parameters for gap opening and
tilt\cite{PhysRevLett.132.096302}. For the latter, by using the numerical
scheme, $V_{0}=0.1$ favors the effective extraction of the spin texture (cf.
circles in Fig. \ref{Qtiltless}(a)) and QGT [light gray lines in Fig.
\ref{Qtiltless}(b)] while $V_{0}=0.001$ works better [black lines for
QGT in Fig. \ref{Qtiltless}(b)] since weak $V_{0}$ implies the accuracy of
Born approximation. If one regards the magnetic impurity as a tip, dual-probe STM should provide an alternative way to characterize the spin
texture\cite{adma.201200257} and then QGT. Furthermore, the lifetime of TISS
can be up to $\sim$ ps \cite{PhysRevLett.115.116801}; this ensures insignificant scattering by other defects, particularly
the nonmagnetic impurities due to backscattering suppression of
TISS\cite{nature08308}. Even considering the finite lifetime of TISS, it
fringes FOs with an exponential decay\cite{PhysRevLett.114.176602} in addition to the
dimension-determined decay $1/R$, which can be well incorporated into our
scheme through the amplitude extraction.

The key of the developed scheme is the information extraction from the
geometric amplitudes of the STM measurement; this can be extended to the other
measurements, e.g., photoemission spectroscopy\cite{RevModPhys.93.025006}.
This information extraction is actually to solve an inverse problem, which
generally requires a numerical treatment, e.g., for the ab initio electronic
structure with an effective two-band description near the Fermi
level\cite{ado6049}, although it is analytical for our considered model
Hamiltonian for gapped TISS. And the related dataset is rather huge, e.g.,
energy-resolved and space-resolved FOs in our scheme. As a result, data
processing technology based on machine learning\cite{RevModPhys.91.045002} is
expected to be developed. In addition, it is interesting to develop the scheme
to locally measure the pseudospin texture\cite{s42005-022-01012-z} and orbital
texture\cite{PhysRevLett.121.086602}, and the relevant QGT.

To conclude, it is a challenging problem to measure the QGT of solid state
systems. To this aim, we develop a spin-polarized STM measurement scheme for
the QGT of two-dimensional solid-state systems, which includes the extraction of
spin texture from geometric amplitudes of FOs induced by the intentionally
introduced magnetic impurity and then the momentum differential of spin
texture for the QGT. Using the TISS as an example, we showcase the application of
the developed scheme. This study theoretically realizes the electric
measurement of the QGT of two-dimensional solid-state systems and highlights the
great potential of STM to obtain the geometric information of electronic structure.

\section*{Acknowledgements}

\noindent{This work was supported by the Innovation Program for Quantum Science and Technology (Grant No. 2024ZD0300104), the Science Challenge Project (Grant No. TZ2025017), the NSFC (Grants No. 12174019, No. 12274019, and No. 12174194), and the NSAF grant in NSFC with Grant No. U2230402. Jia-Ji Zhu is supported by the Natural Science Foundation of Chongqing (Grant No. CSTB2025NSCQ-GPX1272). We acknowledge computational support provided by the Beijing Computational Science Research Center (CSRC) and Hefei Advanced Computing Center.}

\section*{Data availability}

\noindent{The data that support the findings of this article are not publicly available. The data are available from the authors upon reasonable request.}


%

\end{document}